\documentclass{article}
\usepackage{spconf,amsmath,epsfig}
\usepackage{indentfirst}
\usepackage{amssymb}
\usepackage{tabularx}
\usepackage{color}
\usepackage{subfig}
\usepackage{booktabs}
\usepackage{amsthm}
\usepackage{tabularx}
\usepackage{adjustbox}
\usepackage{multicol}
\usepackage{array}
\newcolumntype{L}[1]{>{\raggedright\let\newline\\\arraybackslash\hspace{0pt}}m{#1}}
\newcolumntype{C}[1]{>{\centering\let\newline\\\arraybackslash\hspace{0pt}}m{#1}}
\newcolumntype{R}[1]{>{\raggedleft\let\newline\\\arraybackslash\hspace{0pt}}m{#1}}

\newcommand*{\thead}[1]{%
\multicolumn{1}{c}{\bfseries\begin{tabular}{@{}c@{}}#1\end{tabular}}}
\newcommand{\etal}{\textit{et al.}}

\title{Panicle Counting in UAV Images For \\Estimating Flowering Time in Sorghum}
\name{Enyu Cai\textsuperscript{1}, Sriram Baireddy\textsuperscript{1}, Changye Yang\textsuperscript{1}, Melba Crawford\textsuperscript{2}, and Edward J. Delp\textsuperscript{1}}
\address{\textsuperscript{1}Video and Image Processing Laboratory (VIPER), School of Electrical and Computer Engineering\\
\textsuperscript{2}School of Civil Engineering\\
Purdue University\\
West Lafayette, Indiana, USA}

\begin{document}
%\ninept
%
\maketitle
\begin{abstract}
%Flowering time is the period in which a plant produces flowers. This is typically measured in days from the time the seed was planted. 
Flowering time (time to flower after planting) is important for estimating  plant development and grain yield for many  crops including sorghum.
Flowering time of sorghum can be approximated by counting the number of panicles (clusters of grains on a branch) across multiple dates.
Traditional manual methods for panicle counting are time-consuming and tedious.
In this paper, we propose a method for estimating flowering time and rapidly counting panicles using RGB images acquired by an Unmanned Aerial Vehicle (UAV).
We evaluate three different deep neural network structures for panicle counting and location.
Experimental results demonstrate that our method is able to accurately detect panicles and estimate sorghum flowering time.
\end{abstract}
\section{Introduction}

\begin{figure}[!htbp]
	\centering
	\subfloat[]{\epsfig{figure=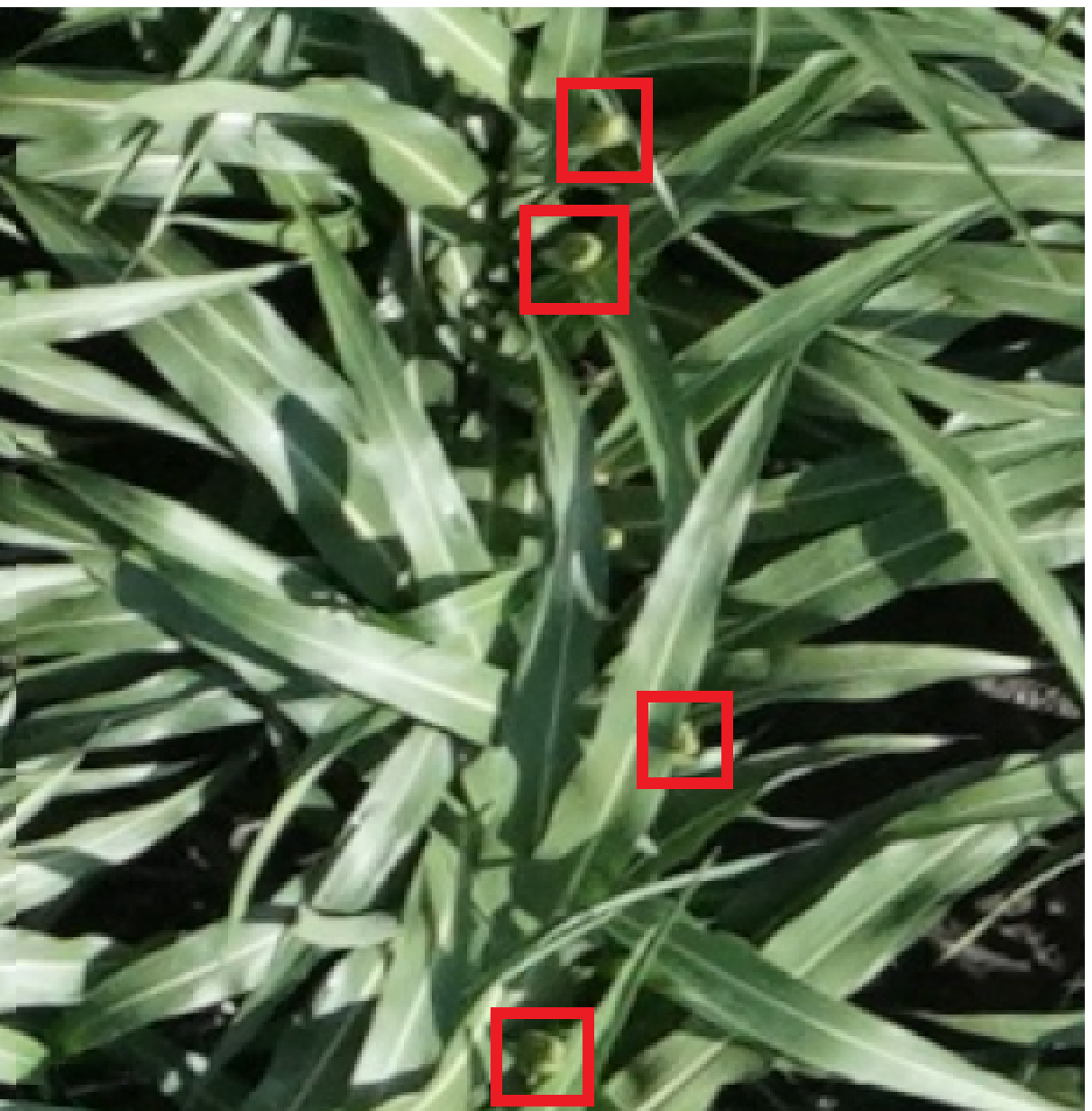,width = 0.19\textwidth}}
	\qquad
	\subfloat[]{\epsfig{figure=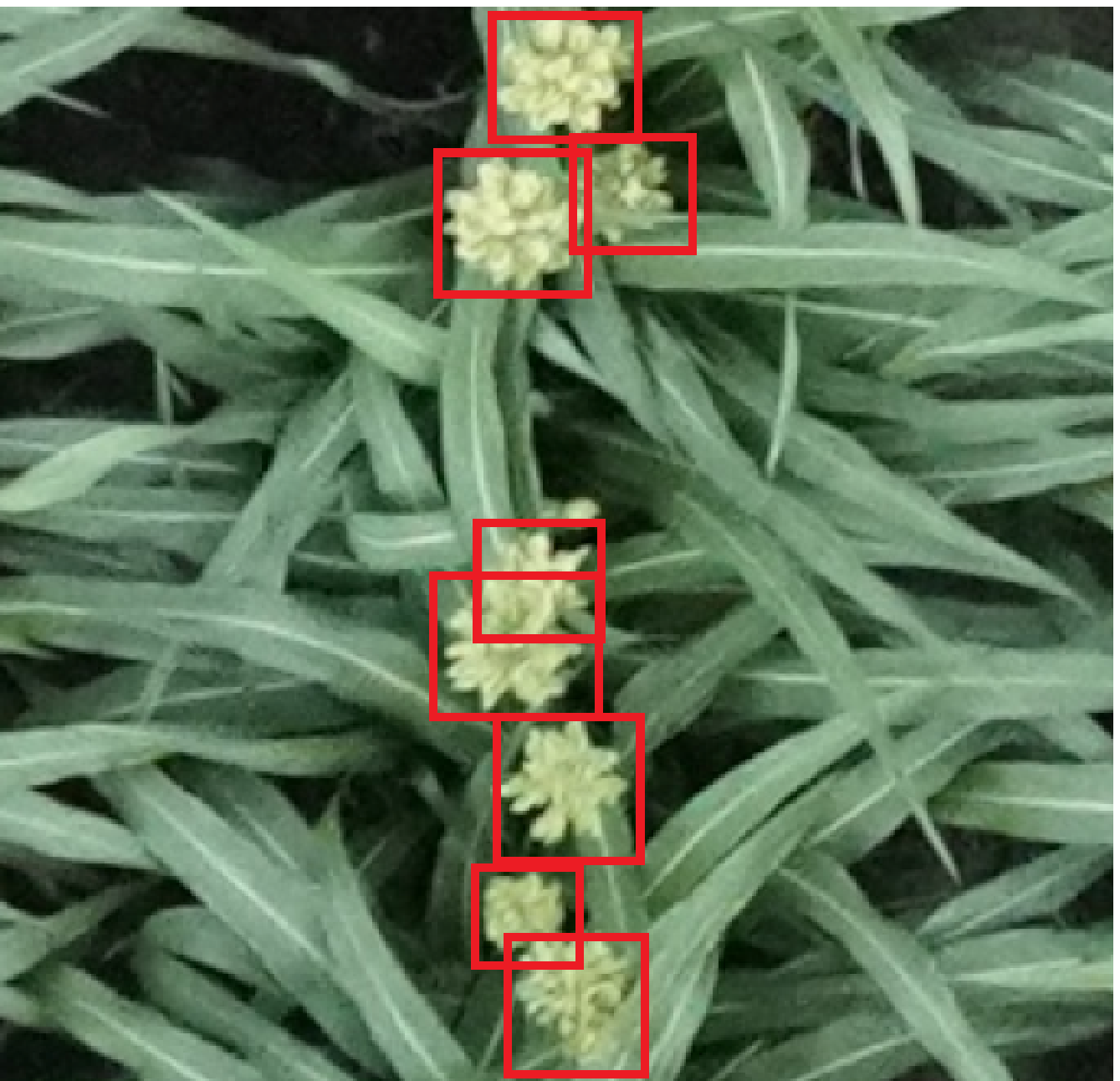,width = 0.20\textwidth}}
	\caption{a) Sorghum plants with panicles labeled using red boxes. In this stage the panicle is not blooming so the plants are not considered as flowering.
	b) Flowering sorghum plants with blooming panicles labeled using red boxes.}
	\label{panicle}
	\vspace*{-0.5cm}
\end{figure}

Sorghum (\textit{Sorghum bicolor} (L.) Moench) is used in biofuels, forage, grain, and food due to its ability to resist water-limited conditions~\cite{sorghum_2010}.
Plant breeders evaluate various properties of a crop during the growing season.
Measurement of physiological properties of plants is known as phenotyping~\cite{walter_2015}.
Flowering time (time to flower after planting) is an important phenotypic trait related to plant development and grain yield in sorghum~\cite{wang_2020}.
A sorghum plant is considered ``flowering'' 
when a panicle (clusters of grains on a branch) is flowering (or blooming), and a plot (a section of the crop field) is flowering when 50\% of the sorghum plants have reached this stage~\cite{sorghum_stage}.
We can evaluate flowering in a sorghum plant by observing its panicles as shown in Figure~\ref{panicle}.
While we are unable to determine the state of flowering of individual panicles due to resolution of most imagery, we can consider counting across temporal data as a potential surrogate measure, as the capability to detect panicles increases when the flowers emerge from the tight panicle.

Traditional phenotyping methods for panicle counting use manual counting, which is time-consuming in large fields with multiple genotypes of plants.
In recent years, the use of Unmanned Aerial Vehicles (UAVs) has been demonstrated for high-throughput phenotyping of many traits~\cite{chapman_2014}.
Compared to traditional phenotyping, UAVs equipped with multiple sensors can collect field data in a non-destructive way and in less time.
For this study, high resolution orthorectified images~\cite{habib} acquired by an RGB camera on a UAV platform were analyzed.
Additional details are included in the description of the datasets below.

Deep neural networks provide promising results for detecting and counting panicles.
In~\cite{ghosal_2019}, Ghosal \etal~developed a weakly supervised deep learning framework with RetinaNet~\cite{retinanet} to detect and count sorghum panicles.
Chandra \etal~proposed an active learning method with Faster-RCNN~\cite{fasterrcnn} for panicle detection in cereal crops~\cite{akshay_2020}.
Segmentation-based networks can be used for panicle detection and counting as well, as shown by Lin \etal~\cite{lin_2020}.
In this paper, we investigate the panicle detection performance of multiple networks and use the counts of the best network for flowering time estimation.

\section{Our Approach}
\label{method}
\begin{figure}[htb]
	\centering
	\centerline{{\epsfig{figure=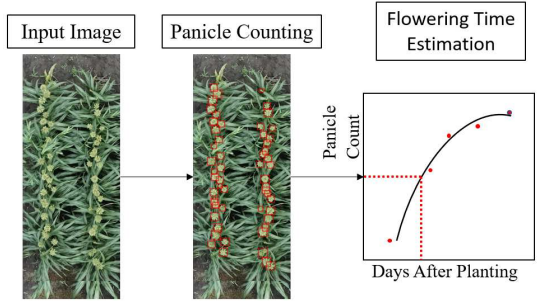,width = 0.4\textwidth}}}
	\caption{Our Approach To Flowering Time Estimation.}
	\label{block}
\end{figure}

Our method consists of multi-temporal panicle detection and flowering time series estimation, as shown in Figure~\ref{block}.
%\textbf{{\color{red} Label the axis in the plot in this figure}}

%\subsection{Datasets}
%\label{subsec:dataset}

For panicle detection training and testing, we use an RGB orthomosaic~\cite{habib} photo of a sorghum field in West Lafayette, Indiana, USA acquired by a Sony ILCE-7RM3 camera mounted on a DJI Matrice 600 Pro platform on July 22, 2020 at 20m altitude.
The orthomosaic photo is cropped into individual images of two row segments of plants.
Each cropped image is horizontally divided into two sub-images.
The images are further separated for training, validation, and testing.
We manually ground truth the images by labeling each panicle with a bounding box.
In total, we have 500 images for training, validation, and testing.
The  images have dimensions of 800 × 600 pixels which are resized to 512 × 512 pixels during training.
Flowering time was estimated for a field of sorghum test plots ({\raise.17ex\hbox{$\scriptstyle\sim$}}200,000 plants/hectare), comprised of two replicates of 80 varieties in a randomized block design (plot size: 7.6m × 3.8m), 10 rows per plot.
%In practice, the flowering time varies for different genotypes of sorghum so the multi-temporal data needs to accommodate different genotypes of sorghum.
In practice, the flowering time varies for different genotypes of sorghum, so this needs to be accounted for.
For this specific genotype, with a planting date of May 13, 2020, we select the multi-temporal RGB images from 65, 68, 70, 76, 79, and 83 days after planting.
Each image is cropped from the associated orthomosaic photo with size of 3000 × 1200 pixels.
The cropped image has 8 row segments of plants because 2 rows in the middle were destructively sampled for biomass.
The ground truth data is obtained by manually counting panicles in these cropped images.

%\subsection{Network Architecture}

We chose the deep networks based on their performance on a general object detection dataset such as COCO~\cite{coco_dataset}.
We selected three detection-based deep networks for panicle detection.

\begin{figure}[htb]
	\centering
	\subfloat[]{\epsfig{figure=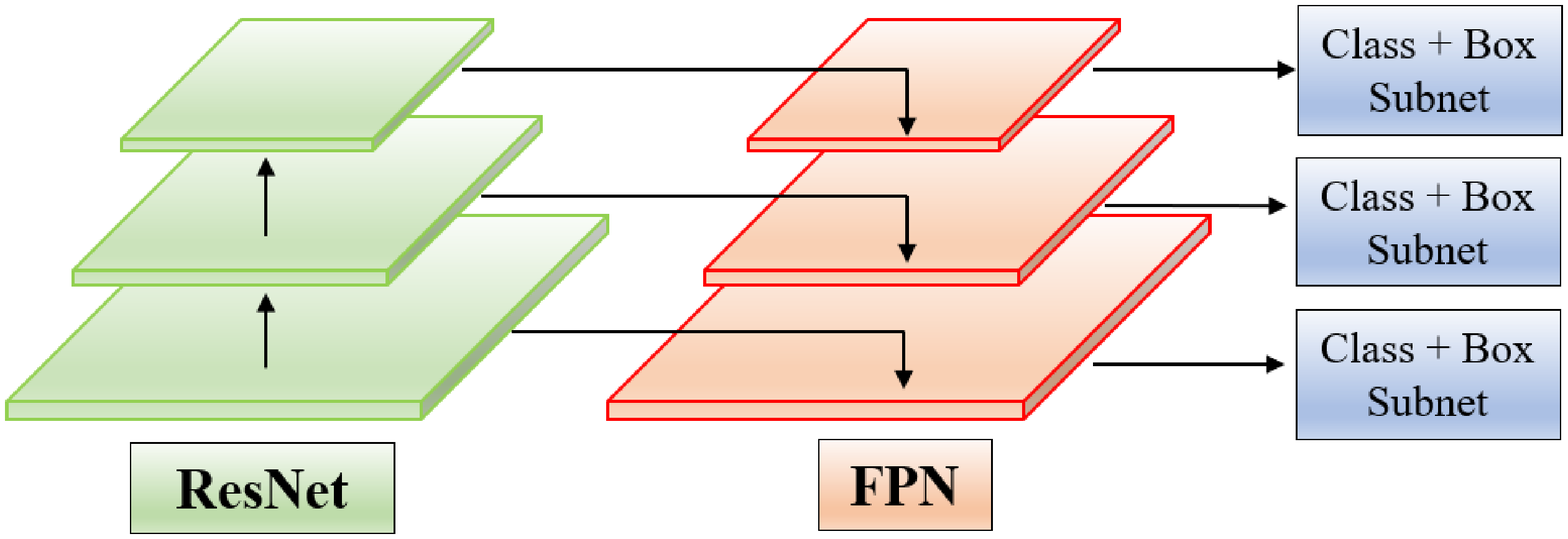,width = 0.25\textwidth}}
	\qquad \qquad \qquad
	\subfloat[]{\epsfig{figure=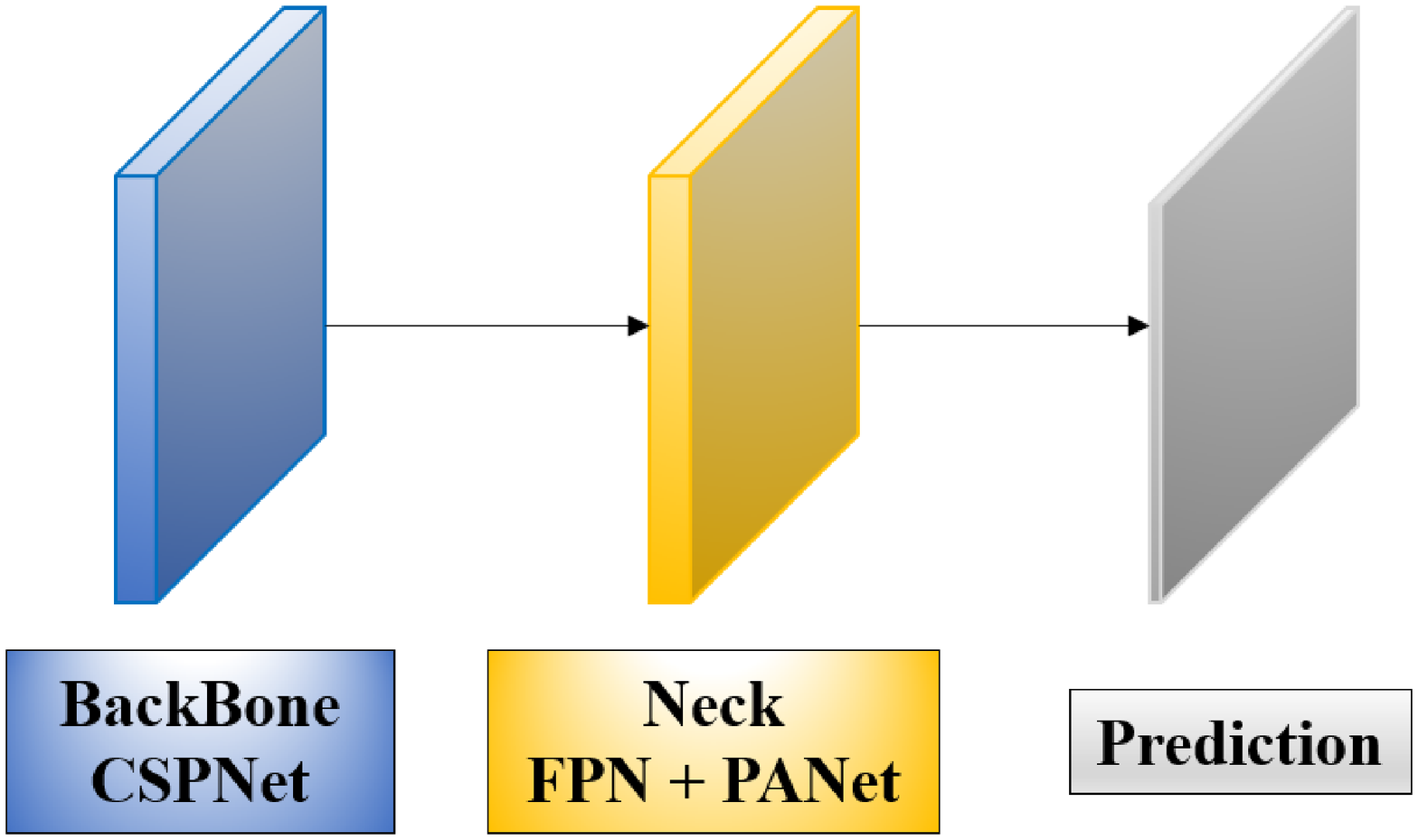,width = 0.2\textwidth}}
	\qquad \qquad \qquad
	\subfloat[]{\epsfig{figure=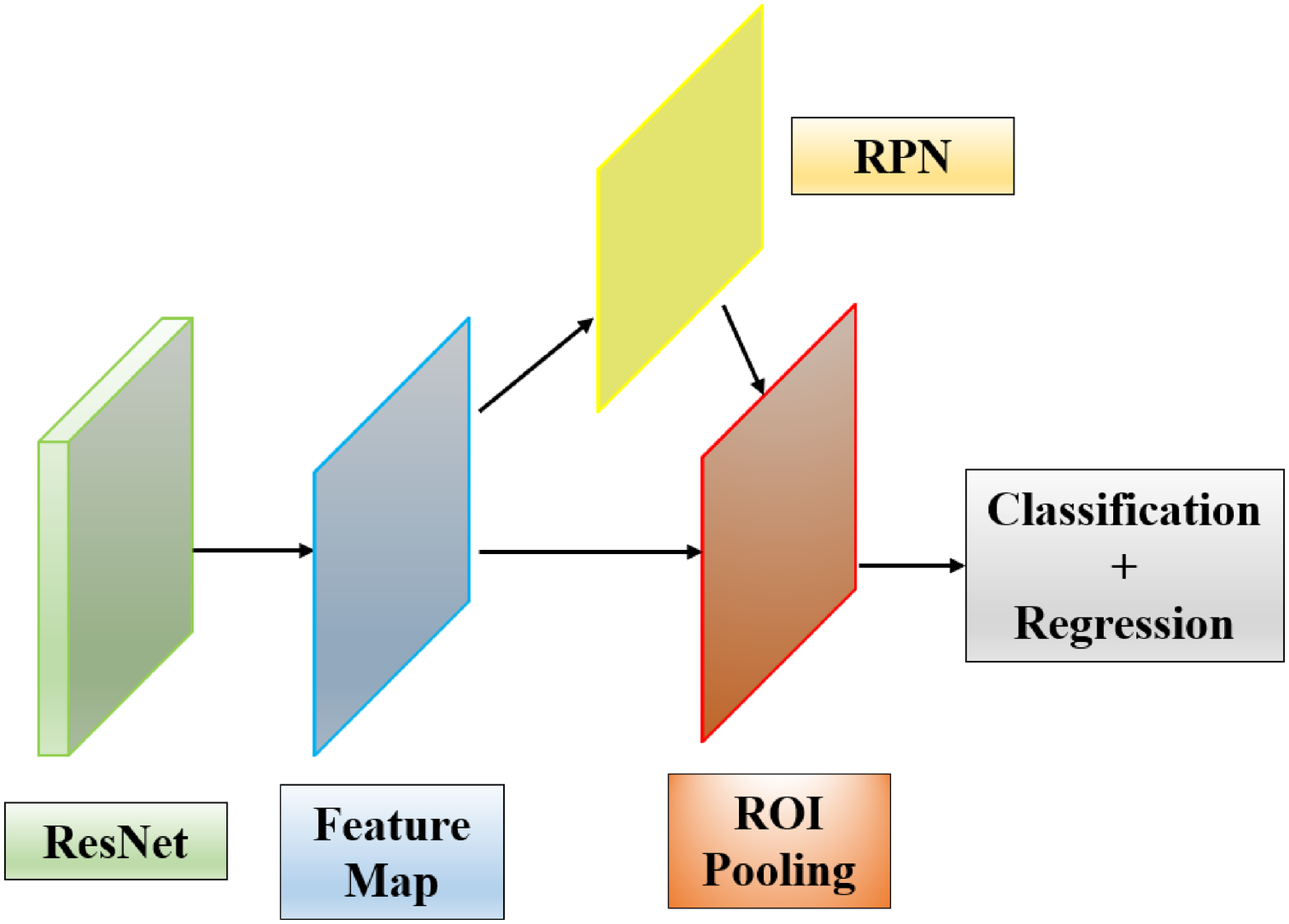,width = 0.25\textwidth}}
	
	\caption{a) RetinaNet.
		     b) YOLOv5.
		     c) Faster-RCNN.}
	\label{network}
\end{figure}

\textbf{RetinaNet.}
RetinaNet~\cite{retinanet} is a one-stage detection-based network with focal loss as the loss function as shown in Figure~\ref{network}.
It uses ResNet~\cite{resnet} and feature pyramid network (FPN)~\cite{fpn} as backbone networks.
Each level of the FPN is connected with a sub-network for bounding box regression and object classification.
The focal loss is used in the classification sub-network.
In our experiments, we choose ResNet-101 with FPN as the backbone for RetinaNet.

\textbf{YOLOv5.}
YOLOv5~\cite{yolov5} is a one-stage detection-based network.
The general structure of YOLOv5 consists of backbone, neck and prediction as shown in Figure~\ref{network}.
YOLOv5 uses CSPNet~\cite{cspnet} as backbone architectures.
FPN~\cite{fpn} and Path Aggregation Network (PANet)~\cite{panet} are used for the neck of YOLOv5.
There are four different versions of YOLOv5.
The main differences of the versions are the depth and width.
We chose the YOLOv5x model for our experiments since it has the best accuracy across the different versions.

\textbf{Faster-RCNN.}
Faster-RCNN~\cite{fasterrcnn} is a two-stage detection-based network consisting of a feature map extractor, regional proposal network (RPN), and Region of Interest (ROI) pooling and classification network as shown in Figure~\ref{network}.
The main idea of Faster-RCNN is to use RPN to generate bounding boxes.
We use the ResNet-101 with FPN as the feature map extractor in the Faster-RCNN model.

\section{Experimental Results}

%\subsection{Panicle Detection and Counting}

We split the 500 images into training (80\%), validation (10\%), and testing (10\%).
For all three networks, we start with models pretrained on the COCO dataset, as this reduces training time.
Learning rate is set to 0.00001 for three networks. 
The training time for each network is around 30 minutes using 4 NVIDIA GTX 1080 Ti graphics cards.
Validation is performed every 10 epochs.

We use Average Precision (AP) with Intersection over Union (IoU) set to 0.5 for panicle detection.
We use Mean Absolute Percent Error (MAPE)~\cite{powers_2011}, Mean Absolute Error (MAE)~\cite{powers_2011}, and Root Mean Squared Error (RMSE)~\cite{powers_2011} for panicle counting.

\begin{multicols}{2}
  \begin{equation}
    \text{Precision (P)} = \frac{\text{TP}}{\text{TP}+\text{FP}}
  \end{equation}\break
  \begin{equation}
    \text{Recall (R)} = \frac{\text{TP}}{\text{TP}+\text{FN}}
  \end{equation}
\end{multicols}

\begin{multicols}{2}
  \begin{equation}
  \label{ap}
    \text{AP} = \sum_{k}(R_{k}-R_{k-1})P_{k}
  \end{equation}\break
  \begin{equation}
  \label{mape}
    \text{MAPE} = \frac{1}{N} \sum_{\substack{i=1}}^{N}\frac{\big| e_i \big|}{C_i}
  \end{equation}
\end{multicols}

\begin{multicols}{2}
  \begin{equation}
  \label{mae}
    \text{MAE} = \frac{1}{N}\sum_{i=1}^{N}| e_i |
  \end{equation}\break
  \begin{equation}
  \label{rmse}
    \text{RMSE} = \sqrt{\frac{1}{N}\sum_{i=1}^{N} \big| e_i \big|^2}
  \end{equation}
\end{multicols}

In these equations, true positive, false positive and false negative are represented by TP, FP and FN, respectively.
In Equation~\ref{ap}, $k$ refers to the $k$-th threshold for precision and recall.
In Equation~\ref{mape},~\ref{mae},~\ref{rmse}, $C_i$ is the ground truth count in the $i$-th image.
$N$ is the number of image samples.

We evaluate the performance of the three networks with the validation and testing datasets.
The results are shown in Table~\ref{tab:validation} and ~\ref{tab:testing}.
Faster-RCNN and YOLOv5 are better than RetinaNet based on four metrics.
YOLOv5 has similar AP and better MAPE, MAE, and RMSE compared to Faster-RCNN.
Based on these results, we use YOLOv5 as the network architecture for flowering time estimation.

\begin{table}[htb]
\centering
\begin{tabular}[c]{cccc}
\toprule
\textbf{Metric}  &  \textbf{RetinaNet}  &  \textbf{YOLOv5}  &  \textbf{Faster-RCNN}  \\\midrule
AP    &  $86.6$         &  $89.1$        &  \textbf{89.8}        \\
MAPE    &  $0.3$          &  \textbf{0.2}        &  $0.3$         \\
MAE     &  $1.8$          &  \textbf{1.2}        &  $1.5$         \\
RMSE    &  $2.5$          &  \textbf{1.8}        &  $2.2$         \\\bottomrule
\end{tabular}
\caption{Evaluation of validation dataset.}
\label{tab:validation}
\end{table}

\begin{table}[htb]
\centering
\begin{tabular}[c]{cccc}
\toprule
\textbf{Metric}  &  \textbf{RetinaNet}  &  \textbf{YOLOv5}  &  \textbf{Faster-RCNN}  \\\midrule
AP     &  $83.8$          &  \textbf{86.2}        &  $86.1$         \\
MAPE    &  $0.2$          &  \textbf{0.1}        &  $0.2$         \\
MAE     &  $3.1$          &  \textbf{1.5}        &  $2.6$         \\
RMSE    &  $4.0$          &  \textbf{2.0}        &  $3.2$         \\\bottomrule
\end{tabular}
\caption{Evaluation of testing dataset.}
\label{tab:testing}
\end{table}

%\subsection{Flowering Time Estimation}

We use a hybrid genotype sorghum with multi-temporal panicle counting ground truth data for flowering time estimation (see Section \ref{method}).
The shape and color of panicles varied for each individual variety of sorghum.
We select the variety based on the similarity of our training data.
We use our panicle counting deep network to estimate the counts for  each test image without resizing.

\begin{figure}[htb]
	\centering
	\centerline{{\epsfig{figure=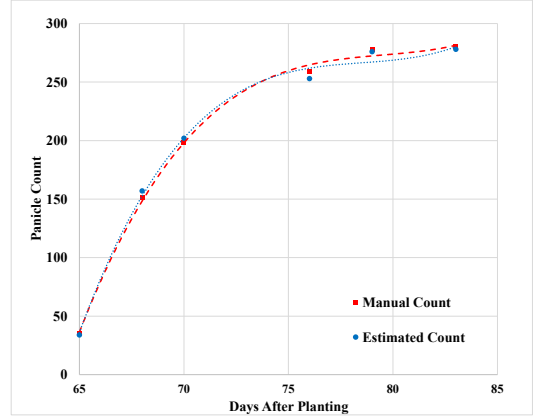,width = 0.39\textwidth}}}
	\caption{Panicle Count Time Series.}
	\label{curve}
\end{figure}

For early dates in the time sequence, some panicles that did not bloom can still be detected by the network.
We set a threshold  for the bounding box size to remove them.
We then fit a third degree polynomial to the estimated counting data to obtain the panicle count time series as shown in Figure~\ref{curve} with the counts in Table~\ref{tab:flowering}.
The estimated flowering time is the intersection between the line associated with half of the ultimate number of panicles counted and the flowering curve.
Our estimated flowering time is 68 days after planting which is nearly identical to the result from the manual counts.

\begin{table}[t]
\centering
\begin{tabular}[c]{ccc}
\toprule
\thead{Days \\ After Planting}  &  \thead{Manual \\ Count}  &  \thead{Estimated \\ Count} \\ \midrule
65     &  35          &  34  \\
\thead{68 \\ (Est. Flowering Time)}    &  151          &  157 \\
70     &  198          & 202 \\
76     &  259          & 253 \\
79     &  278          & 276 \\
83    &  280          &  278 \\
\bottomrule
\end{tabular}
\caption{Flowering time estimation.}
\label{tab:flowering}
\end{table}

\section{Conclusion and Discussion}

In this paper, we propose a method for flowering time estimation by counting panicles in UAV images.
We evaluate the performance of three popular detection-based network architectures and show that YOLOv5 has the best performance.
We also describe the use of multi-temporal panicle counting for flowering time estimation.
Our result shows the estimated flowering times are nearly identical to the results of manual counting.
Future work will include training with panicle images with different shape and color to generalize the method for more varieties of sorghum plants.

\section{Acknowledgments}

We thank Professor Ayman Habib and the Digital Photogrammetry Research Group (DPRG) from the School of Civil Engineering at Purdue University for providing the images used in this paper. The work presented herein was funded in part by the Advanced Research Projects Agency-Energy (ARPA-E), U.S. Department of Energy, under Award Number DE-AR0001135.
The views and opinions of the authors expressed herein do not necessarily state or reflect those of the United States Government or any agency thereof.
Address all correspondence to Edward J. Delp, ace@ecn.purdue.edu

\bibliographystyle{IEEEbib}
\bibliography{ref}

\end{document}